\documentclass[aps,twocolumn,showpacs]{revtex4}
\usepackage{mathrsfs}
\begin{document}
\title{Spin and orbital angular momentum in gauge theories (II):
QCD and nucleon spin structure}
\author{Xiang-Song Chen\footnote{Email: cxs@scu.edu.cn}, Xiao-Fu L\"{u}}
\affiliation{Department of Physics, Sichuan University, Chengdu
610064, China}
\author{Wei-Min Sun, Fan Wang}
\affiliation{Department of Physics, Nanjing University, Nanjing
210093, China}
\author{T. Goldman}
\affiliation{Theoretical Division, Los Alamos National Laboratory,
Los Alamos, NM 87545, USA}
\date{\today}

\begin{abstract}
Parallel to the construction of gauge invariant spin and orbital
angular momentum for QED in paper (I) of this series \cite{Chen07},
we present here an analogous but non-trivial solution for QCD.
Explicitly gauge invariant spin and orbital angular momentum
operators of quarks and gluons are obtained. This was previously
thought to be an impossible task, and opens a more promising avenue
towards the understanding of the nucleon spin structure.
\pacs{11.15.-q, 13.88.+e, 14.20.Dh,14.70.Dj}
\end{abstract}
\maketitle

 As a composite particle, the nucleon naturally gets its spin from
the spin and orbital motion of its constituents: quarks and gluons.
From a theoretical point of view, the first task in studying the
nucleon spin structure is to find out the appropriate operators for
the spin and orbital angular momentum of the quark and gluon fields.
Given these operators, one can then study their matrix elements in a
polarized nucleon state, and investigate how these matrix elements
can be related to experimental measurements. Pitifully and
surprisingly, after 20 years of extensive discussions of the nucleon
spin structure \cite{Bass07,Herm07,Stra07,Bass05}, this first task
was never done, and even largely eluded the attention of the
community.

At first thought, it seems an elementary exercise to derive the
quark and gluon angular momentum operators. From the QCD Lagrangian
\begin{equation}
\mathscr{L}=-\frac 14F^a_{\mu\nu}F^{a\mu\nu}+ \bar\psi\left(
i\gamma^\mu D_\mu -m \right)\psi ,\label{L}
\end{equation}
one can promptly follow N\"{o}ther's theorem to write down the
conserved QCD angular momentum:
\begin{eqnarray}
\vec J_{QCD} &=& \int d^3 x \psi ^\dagger \frac 12 \vec \Sigma \psi
+ \int d^3x \psi ^\dagger \vec x \times\frac 1i \vec \nabla \psi
\nonumber \\
 &+&\int d^3x \vec E^a\times \vec A^a+
\int d^3x E^{ai}\vec x\times \vec \nabla A^{ai} \nonumber \\
&\equiv& \vec S_q +\vec L_q +\vec S_g +\vec L_g, \label{J1}
\end{eqnarray}
and readily identify the four terms here as the quark spin, quark
orbital angular momentum, gluon spin, and gluon orbital angular
momentum, respectively. However, except for the quark spin, all
other three terms are gauge dependent, therefore have obscure
physical meanings. In this regard, it should be noted that the total
$J_{QCD}$ is still gauge invariant (as it should be). This can be
seen from an alternative, explicitly gauge invariant expression
\begin{eqnarray}
\vec J_{QCD} &=& \int d^3 x \psi ^\dagger \frac 12 \vec \Sigma \psi
+ \int d^3x \psi ^\dagger \vec x \times\frac 1i \vec D
\psi\nonumber\\
&+&\int d^3x \vec x\times \left(\vec E^a\times\vec B^a\right) \nonumber \\
&\equiv& \vec S_q +\vec L'_q +\vec J'_g.  \label{J2}
\end{eqnarray}
It is obtained from Eq. (\ref{J1}) by adding a surface term which
vanishes after integration. Since all terms in Eq. (\ref{J2}) are
separately gauge invariant, one may intend to identify $\vec L'_q$
as the quark orbital angular momentum, and $J'_g$ as the total gluon
angular momentum. However, a further decomposition of $\vec J'_g$
into gauge invariant gluon spin and orbital parts is lacking.
Moreover, $\vec L'_q$ and $\vec J'_g$ do not obey the fundamental
angular momentum algebra, $\vec J \times \vec J =i \vec J$, hence
they are not the relevant rotation generators. It has been long
assumed by the community that the reconciliation of gauge invariance
and angular momentum algebra is not possible, and that gauge
invariant local gluon spin and orbital angular momentum operators do
not exist. \cite{Bass05}

Since QED is also a gauge theory, the above problems already emerged
there. \cite{Calv06,Enk94} Actually, by dropping the color indices,
Eqs. (\ref{J1}) and (\ref{J2}) become exactly the expressions for
the QED angular momentum. In the first paper of this series
\cite{Chen07}, we have provided a satisfactory and decisive answer
for the spin and orbital angular momentum in QED:
\begin{eqnarray}
\vec J_{QED} &=& \int d^3 x \psi ^\dagger \frac 12 \vec \Sigma \psi
+ \int d^3x \psi ^\dagger \vec x \times\frac 1i \vec D_{pure} \psi
\nonumber \\
 &+&\int d^3x \vec E \times \vec A_{phys}+ \int d^3x
E^i\vec
x\times \vec \nabla A_{phys}^i \nonumber \\
&\equiv& \vec S_e +\vec L_e +\vec S_\gamma +\vec L_\gamma
\label{QED}.
\end{eqnarray}
Here, $\vec D_{pure} \equiv\vec \nabla -ie\vec A_{pure}$, $\vec
A_{pure}+\vec A_{phys}\equiv \vec A$ are defined through:
\begin{eqnarray}
\vec \nabla\cdot \vec A_{phys} =0,\label{A2}\\
\vec \nabla \times \vec A_{pure}=0. \label{A3}
\end{eqnarray}
Eq. (\ref{A3}) tells that $\vec A_{pure}$ is a pure gauge field in
QED. With the boundary condition that $\vec A$, $\vec A_{pure}$, and
$\vec A_{phys}$ all vanish at spatial infinity, Eqs. (\ref{A2}) and
(\ref{A3}) prescribe a unique decomposition of $\vec A$ into $\vec
A_{pure}$ and $\vec A_{phys}$, and dictates their gauge
transformation properties: The pure gauge field $\vec A_{pure}$
transforms in the same manner as the full vector field $\vec A$
does, while $\vec A_{phys}$ is gauge invariant, thus can be regarded
as the ``physical'' part of $\vec A$.

Eq. (\ref{QED}) qualifies for the correct expressions of spin and
orbital angular momentum of electrons and photons in that:
\cite{Chen07}
\begin{enumerate}
\item each term in Eq. (\ref{QED}) is separately gauge invariant;
\item each term in Eq. (\ref{QED}) is the correct generator for
relevant rotations;
\item $\vec S_\gamma$ and $\vec L_\gamma$ give correct result for
the angular distribution of angular momentum flow in polarized
atomic radiations, whereas the took-for-granted expression $\vec
x\times \left(\vec E\times \vec B\right)$ {\em does not}.
\end{enumerate}

In fact, Eq. (\ref{QED}) is also physically more reasonable than
Eqs. (\ref{J1}) and (\ref{J2}): the photon angular momentum should
contain only the ``physical'' part of the gauge field, which
nevertheless should not appear in the orbital angular momentum of
the electron. The latter should thus only include the non-physical
$A_{pure}$ to cancel the also non-physical phase dependence of the
electron field, keeping the whole $\vec L_e$ gauge invariant. From
these considerations, it is natural to expect that the correct,
gauge invariant expression of QCD angular momentum should be:
\begin{eqnarray}
\vec J_{QCD} &=& \int d^3 x \psi ^\dagger \frac 12 \vec \Sigma \psi
+ \int d^3x \psi ^\dagger \vec x \times\frac 1i \vec D_{pure} \psi
\nonumber \\
 &+& \int d^3x \vec E^a \times \vec A^a_{phys}+ \int
d^3x
E^{ai}\vec x\times \vec \nabla A^{ai}_{phys} \nonumber \\
&\equiv& \vec S_q +\vec L''_q +\vec S''_g +\vec L''_g \label{J3},
\end{eqnarray}
where $\vec D_{pure} \equiv\vec \nabla -ig\vec A_{pure}$ and $\vec
A_{pure}\equiv T^a\vec A^a_{pure}$. The essential task now is to
properly define the physical field $\vec A^a_{phys}$ and the pure
gauge field $\vec A^a_{pure}$ so that they have the desired gauge
transformation properties, and to prove that the sum of the four
terms in Eq. (\ref{J3}) equals that in Eqs. (\ref{J1}) and
(\ref{J2}). This turns out to be non-trivial.

The construction of Eqs. (\ref{A2}) and (\ref{A3}) obviously does
not work in QCD: For one thing, $\vec A_{pure}$ defined by Eq.
(\ref{A2}) is not a pure gauge term in QCD; for another thing, Eqs.
(\ref{A2}) and (\ref{A3}) are not invariant under the SU(3) gauge
transformation
\begin{equation}
A_\mu\rightarrow A'_\mu =UA_\mu U^\dagger -\frac ig U\partial_\mu
U^\dagger.
\end{equation}

To make $\vec A_{pure}$ a pure gauge term in QCD, we should require
instead of Eq. (\ref{A3}):
\begin{equation}
\vec D_{pure}\times \vec A_{pure} =\vec \nabla \times \vec A_{pure}
-ig \vec A_{pure}\times \vec A_{pure}=0. \label{def1}
\end{equation}
This provides two independent equations for $\vec A_{pure}$. We
still need a third equation playing the same role as Eq. (\ref{A2})
does in QED, so that $\vec A^a_{phys}$ and $\vec A^a_{pure}$ have
the desired transformation:
\begin{eqnarray}
\vec A_{pure}&\rightarrow & \vec A'_{pure}= U\vec A_{pure} U^\dagger
+\frac ig U\vec
\nabla U^\dagger, \label{Apure}\\
\vec A_{phys}&\rightarrow &\vec A'_{phys}= U\vec A_{phys} U^\dagger.
\label{Aphys}
\end{eqnarray}

To seek this third equation, we go inversely by applying these
transformation to examine the gauge invariance of the angular
momentum operators. (The reason why this is possible will be clear
shortly below.)

The quark orbital angular momentum $\vec L''_q$ provides no further
constraints. Eqs. (\ref{def1}) and (\ref{Apure}) guarantee its gauge
invariance, as well as the correct algebra $\vec L''_q\times \vec
L''_q=i\vec L''_q$.

The gluon spin $\vec S''_g$ provides no further constraints either.
Eq. (\ref{Aphys}) tells that it is gauge invariant.

The situation for the gluon orbital angular momentum $\vec L''_g$ is
different: Unlike in QED, $\vec A_{phys}$ here is gauge covariant
instead of invariant, which leads to the gauge transformation of
$\vec L''_g$:
\begin{eqnarray}
&&E_a^i\vec x\times \vec \nabla
A^{ai}_{phys}=2\textrm{Tr}\left\{E^i\vec x\times \vec \nabla
A_{phys}^i \right\}\nonumber \\ &\rightarrow&
2\textrm{Tr}\left\{UE^iU^\dagger\vec x\times \vec \nabla \left(U
A_{phys}^i U^\dagger
\right)\right\}\nonumber \\
&=& 2\textrm{Tr}\left\{E^i\vec x\times \vec \nabla A_{phys}^i
\right\} \nonumber \\
& +& 2 \textrm{Tr}\left\{\vec x\times U^\dagger \left(\vec\nabla
U\right)\left(\vec A_{phys}\cdot \vec E-\vec E\cdot\vec
A_{phys}\right) \right\}
\end{eqnarray}
Now, to make $\vec L''_g$ invariant under arbitrary gauge
transformations, we have to set
\begin{equation}
\left[\vec A_{phys}, \vec E \right]\equiv\vec A_{phys}\cdot \vec
E-\vec E\cdot\vec A_{phys}=0. \label{def2}
\end{equation}
This is the third equation we are seeking. The remaining task is to
make a consistency cross-check of whether Eqs. (\ref{def1}) and
(\ref{def2}) dictate the transformation properties in  Eqs.
(\ref{Apure}) and (\ref{Aphys}).

Before this cross-check, we first make another vital check, namely,
whether the definitions of $\vec A_{pure}$ and $\vec A_{phys}$ by
Eqs. (\ref{def1}) and (\ref{def2}) would let the total angular
momentum in Eq. (\ref{J3}) equal that in Eqs. (\ref{J1}) and
(\ref{J2}). Since no more tricks can be played, we can only pray for
a positive answer. A slightly lengthy but straightforward
calculation shows that the answer is indeed positive.

As to the cross-check, we note that $\vec A'_{pure}$ and $\vec
A'_{phys}$ given by Eqs. (\ref{Apure}) and (\ref{Aphys}) are
solutions of
\begin{eqnarray}
\vec D'_{pure}\times \vec A'_{pure}=0, \label{def1'}\\
\left[\vec A'_{phys}, \vec E' \right]=0, \label{def2'}
\end{eqnarray}
where $\vec E'=U\vec E U^\dagger$. The question is whether Eqs.
(\ref{def1'}) and (\ref{def2'}) have solutions other than that given
by Eqs. (\ref{Apure}) and (\ref{Aphys}). Equivalently, this is to
say whether Eqs. (\ref{def1}) and (\ref{def2}) uniquely determine
the decomposition of $\vec A$ into $\vec A_{pure}$ and $\vec
A_{phys}$, or, essentially, whether the constraint $\left[\vec A,
\vec E \right]=0$ would completely fix the gauge. This is a tricky
question, for, unlike in QED, many gauges in QCD suffer from
topological complexity such as Gribov copies \cite{Grib78}.
Fortunately, such complexity does not bother us here: If
supplementary conditions are needed to restrict the solutions of
Eqs. (\ref{def1'}) and (\ref{def2'}) to that given by Eqs.
(\ref{Apure}) and (\ref{Aphys}), they can be simply added, without
affecting the equality of Eq. (\ref{J3}) with Eqs. (\ref{J1}) and
(\ref{J2}), and without affecting the gauge invariance of the
angular momentum operators we constructed.

We close this paper with the following remarks:
\begin{enumerate}
\item For QED in the Coulomb gauge $\vec \nabla \cdot \vec A=0$,
Eq. (\ref{QED}) coincides with Eq. (\ref{J1}) (with color indices
dropped). Similarly, for QCD in the gauge $\left[\vec A, \vec E
\right]=0$ (together with possible supplementary conditions to
completely fix the gauge), Eq. (\ref{J3}) coincides with Eq.
(\ref{J1}). The gauge $\left[\vec A, \vec E \right]=0$ has the sense
of a ``generalized'' Coulomb gauge, for it leads to the equation of
motion: $\vec \nabla \cdot \vec E^a=g\psi^\dagger T^a\psi$, similar
to the Gauss law in QED.
\item Both in QED and QCD, the decomposition of $\vec A$ into $\vec
A_{pure}$ and $\vec A_{phys}$ is not Lorentz invariant. This means
that whenever a Lorentz boost is made, the decomposition has to be
redone, and the angular momentum operators have to be redefined
accordingly. This brings no essential trouble, because the angular
momentum operators are not Lorentz invariant anyhow, and an
experimental observer {\em knows} which reference frame an object is
in. This is in contrast to gauge invariance requirement, because an
observer {\em can never know} which gauge an object is in.
\item The so-called gluon polarization $\Delta G$ being measured at
several facilities \cite{Stra07} is related to $\vec S_g$ in Eq.
(\ref{J1}) in the temporal gauge in the infinite momentum frame of
the proton. \cite{Jaff96} From our discussion, $\Delta G$ is not the
gauge invariant gluon spin $S''_g$ we constructed here.
\item Direct measurement of the photon spin has been performed by
Beth over 70 years ago. \cite{Beth36} Detection and manipulation of
the photon orbital angular momentum have also been carried out
recently, and became a hot topic due to its potential application in
quantum information processing.
\cite{Enk07,Ande06,Alex06,Marr06,Leac02,Alle92} These measurements
can be perfectly interpreted with the operators in Eq. (\ref{QED}).
\cite{Chen07,Calv06,Enk94} This encourages people to investigate the
picture of the nucleon spin in terms of the gauge-invariant,
physically meaningful decomposition in Eq. (\ref{J3}), which is
exactly analogous to Eq. (\ref{QED}) for QED. Experimentally, the
free-beam-based photon measurements can certainly not be extended to
gluons, and appropriate (probably ingenious) methods for measuring
$L''_q$, $S''_g$, and $L''_g$ have to be invented; but the explicit
gauge invariance of these quantities guarantee at least pertinent
theoretical calculations of them, especially in lattice QCD.
\end{enumerate}

The authors acknowledge a helpful discussion with C.D. Roberts. This
research is supported in part by the China National Science
Foundation under grants 10475057 and 90503011, and in part by the
U.S. Department of Energy under contract W-7405-ENG-36.

\end{document}